# Pressure induced band structure evolution of halide perovskites: a first-principles atomic and electronic structure study

Yang Huang[1,2], Lingrui Wang[3], Zhuang Ma[1], and Fei Wang[1,*]

[1] International Laboratory for Quantum Functional Materials of Henan, School of Physics and Engineering, Zhengzhou University, Zhengzhou 450001, China

[2] Materials Science and Engineering Program, University of California San Diego, La Jolla, California 92093, USA

[3] Key Laboratory of Materials Physics of Ministry of Education, School of Physics and Engineering, Zhengzhou University, Zhengzhou 450001, China

## Abstract

Density functional theory (DFT) based calculations have been conducted to draw a broad picture of pressure induced band structure evolution in various phases of organic and inorganic halide perovskite materials. Under a wide range of pressure applied, distinct band structure behaviors including magnitude change of band gap, direct-indirect/indirect-direct band gap transitions and CBM/VBM shifts, have been observed between organic and inorganic perovskites among different phases. Through atomic and electronic structure calculations, band gap narrowing/widening has been rationalized through crystal orbitals coupling transformations; direct-indirect mutual transitions were explained based on structural symmetry evolution; different VBM/CBM shifts behaviors between organic and inorganic perovskites were analyzed focusing on orientation and polarity of molecules/atoms outside the octahedrals. These results provide a comprehensive guidance for further experimental investigations on pressure engineering of perovskite materials.

[*] E-mail: wfei@zzu.edu.cn

# Introduction

Halide perovskites have emerged as promising materials for optoelectronic devices and solar cells. The power conversion efficiency of solar cells based on this family of materials has significantly increased in the past several years, from 3.8% [1] to 23.3% [2]. Halide perovskites can be typically divided into four classes: organic lead perovskite, inorganic lead perovskite, organic lead-free perovskite and inorganic lead-free perovskite, each of which owns outstanding advantages. For example, $MAPbX_3$ (MA= methylammonium; X=Cl, Br, I) stands out in low manufacturing cost [3], $CsPbX_3$ (X=Cl, Br, I) performs higher thermal stability than its organic counterparts [4], and lead-free ones get rid of toxicity which impedes the commercialization of perovskites.

Band gap of perovskite, the proxy for evaluating the ability to absorb light in solar spectrum, is able to be tunned in three ways: temperature induced phase transition (typically, Eg,cubic < Eg,tetragonal < Eg, orthorhombic [5-6]), chemical modification [7], and hydrostatic pressure. Compared to others, pressure is a green and facile method that can successively alter band gap. Kong *et al*. provided experimental evidence that by applying hydrostatic pressure to $MAPbI_3$ and $MAPbBr_3$, band gap narrows down and then widens up with increasing pressure [8]. Xiao *et al*. experimentally discovered that increasing the pressure applied on orthorhombic nanocrystal $CsPbBr_3$ can sequentially induce band gap narrowing, widening and direct-to-indirect transformation [9]. Ying *et al*. theoretically investigated pressure effects on various cubic inorganic halide perovskites through different computational methods and observed that band gap narrows down to zero and then inverses as pressure increases [10]. According to these reported results, it seems that band gap under pressure behaves differently in various perovskite crystal structures (organic vs inorganic and different phases), so it is necessary to perform a comparative study among different phases and between organic and inorganic crystals. For different phases, the intrinsic difference is structural symmetry. In a three dimentional perspective, for inorganic halide perovskite, cubic has the highest structural symmetry, tetragonal is intermediate and orthorhombic phases have the lowest symmetry. For organic halide perovskites, due to the orientation and polarity of MA molecules, organic halide perovskites often show lower symmetry compared with their inorganic counterparts.

In this work, for representativity, highly symmetric cubic CsPbBr$_3$ and cubic CsSnBr$_3$, less symmetric cubic MAPbBr$_3$, cubic MASnBr$_3$, orthorhombic CsPbBr$_3$ and orthorhombic CsSnBr$_3$ are selected and by applying the first-principles calculations, evolutions of band gap properties under a wide range of pressure are systematically investigated. Under a wide range of pressure applied, distinct band structure behaviors, from semiconductor developed to metilic by pressure is studied.

Computational methods

The first-principles calculations have been performed using the plane-wave pseudopotential as implemented in the VASP code [11-12]. The electron-core interactions were described with the frozen-core projected augmented wave pseudopotentials [13]. The generalized gradient approximation formulated by Perdew, Burke, and Ernzerhof (PBE) as the exchange-correlation functional [14] with cut-off energy of 500 eV for organic and 300 eV for inorganic halide perovskites was chosen in all of our calculations. A reciprocal space sampling with R center of 7 × 7 × 7 and Γ center of 5 × 3 × 5 Monkhorst-Pack mesh [15] of the Brillouin zone was used in the cubic and orthorhombic structural optimizations. All the structures are fully relaxed until the forces on each atom are smaller than 0.01 eV/Å with a tetrahedron method with Blöchl corrections in broadening of 0.05 eV. According to our test, MA molecules in cubic MAPbBr$_3$ and cubic MASnBr$_3$ tend to orient along <111> direction due to lower energy compared to <110> and <100> directions. This is consistent with the results on MAPbI$_3$ by Giorgi et al. [16]. Therefore, MA orientation was set to be <111> direction in all of the calculation. Pressure was modeled with fixing volume while allowing all lattice constants and atomic positions to relax. Pressure values were calculated through the relation $p = -\frac{\partial E}{\partial V}$, where E is the energy of the crystal and V is the cell volume.

As has been documented, DFT-PBE shows an obvious underestimation on band gap calculation relative to experimental measurements. An advanced computationally expensive method, spin-orbital coupling GW quasiparticle correlation (soc-GW), has been proved to be accurate in band gap calculations [17]. However, DFT-PBE is still effective for the discussions in this work. It is because that it has been shown that compared with band structure calculated by DFT-PBE, soc-GW changes neither band edge orbital characters nor band gap position in k-space [17-18], thus DFT-PBE is able to qualitatively provide the picture of band gap properties evolution under pressure.

Results and discussion

The six perovskite structures studied in this work with both calculated and reported lattice constants and band gap under ambient pressure are listed in Table 1. As has been demonstrated experimentally and theoretically, these unpressed perovskites are direct band gap semiconductors.

|  | Calculated results in this work | | | | Reported results | | | |
|---|---|---|---|---|---|---|---|---|
|  | a (Å) | b (Å) | c (Å) | Eg (eV) | a (Å) | b (Å) | c (Å) | Eg (eV) |
| Cubic $MAPbBr_3$ | 6.0926 | —— | —— | 1.9761 | 5.9291[8] | —— | —— | 1.971[19], 1.96[20], 2.18[21] |
| Cubic $MASnBr_3$ | 6.1296 | —— | —— | 1.4628 | 5.97[21] | —— | —— | 1.122[22] |
| Cubic $CsPbBr_3$ | 5.9924 | —— | —— | 1.7716 | 5.84[22] | —— | —— | 1.6[23] |
| Cubic $CsSnBr_3$ | 5.8880 | —— | —— | 0.6334 | 5.887[23] | —— | —— | 0.627[24] |
| Orthorhombic $CsPbBr_3$ | 8.5241 | 11.9125 | 8.2162 | 2.1163 | 8.207[25] | 8.255[25] | 11.759[25] | 2.14 [9], 2.30 [26] |
| Orthorhombic $CsSnBr_3$ | 8.3634 | 11.7609 | 8.1782 | 0.9442 | 8.1965[27] | 11.5830[27] | 8.0243[27] | 1.01[28] |

Table 1: DFT-PBE calculated and reported lattice constants and band gap values of six crystal structures studied in this work.

Band gap value variation of all six crystals under pressure is computed and plotted in Figure 1.

The band edge gap represents the energy difference between conduction band edge (i.e. CBM) and valence band edge (i.e. VBM) in the total density of states while the R-R ($\Gamma$-$\Gamma$) band gap equals to the energy difference between conduction band and valence band at R ($\Gamma$) point, the Brillouin zone position under ambient pressure. According to Figure 1, for the two crystal structures with the highest structural symmetry, cubic $CsPbBr_3$ and $CsSnBr_3$ (Figure 1 (a) (b)), band gap remains direct and monochromatically decreases down to zero as pressure increases, which is consistent with previous computational result [10]. For the other four less symmetric structures, cubic $MAPbBr_3$, cubic $MASnBr_3$, orthorhombic $CsPbBr_3$ and orthorhombic $CsSnBr_3$, interesting variations on band gap occur. According to Figure 1 (c)-(f), four critical pressures (CP_I-IV) exist: (1) Under CP_I (0~4 GPa), all crystals have direct band gap and show band gap contraction with increasing pressure; (2) When pressure reaches beyond CP_I, band gap widening happens. Band gap of organic perovskites becomes indirect while band gap of inorganic perovskites remains direct; (3) With pressure being continuously increased beyond CP_II (around 5 GPa), inorganic perovskites have their band gap become indirect; (4) As pressure keeps going up to CP_III (8~15 GPa), bandgaps of all crystals begin to decrease; (5) When pressure goes beyond CP_IV (10~25 GPa), band gap of inorganic perovskites turn back to direct again while organic perovskites still maintain in indirect band gap; (6) Bandgaps of all crystals will decrease to zero without any direct-indirect/indirect-direct band gap transition under further increasing pressure. (1) (2) and (3) are consistent with previous experimental results on cubic $MAPbBr_3$ [8] and orthorhombic $CsPbBr_3$ [9]; (4) and (6) are consistent with previous experimental result on cubic $MAPbI_3$ [29]; (5) has no reported correspondings. Band structure close to the Fermi level of cubic $MAPbBr_3$ and orthorhombic $CsPbBr_3$ (similar for $MASnBr_3$ and $CsSnBr_3$ and thus not presented) under four critical pressure values are plotted in Figure 2. According to Figure 2, when direct-to-indirect and indirect-to-direct band gap alternation happens, in organic perovskites, both conduction band minimum (CBM) and valence band maximum (VBM) move away from the original R point in Brillioun zone; while in inorganic perovskites, only VBM moves away from the original Γ point and CBM remains unmoved. These two revealed trends have no reported correspondings either.

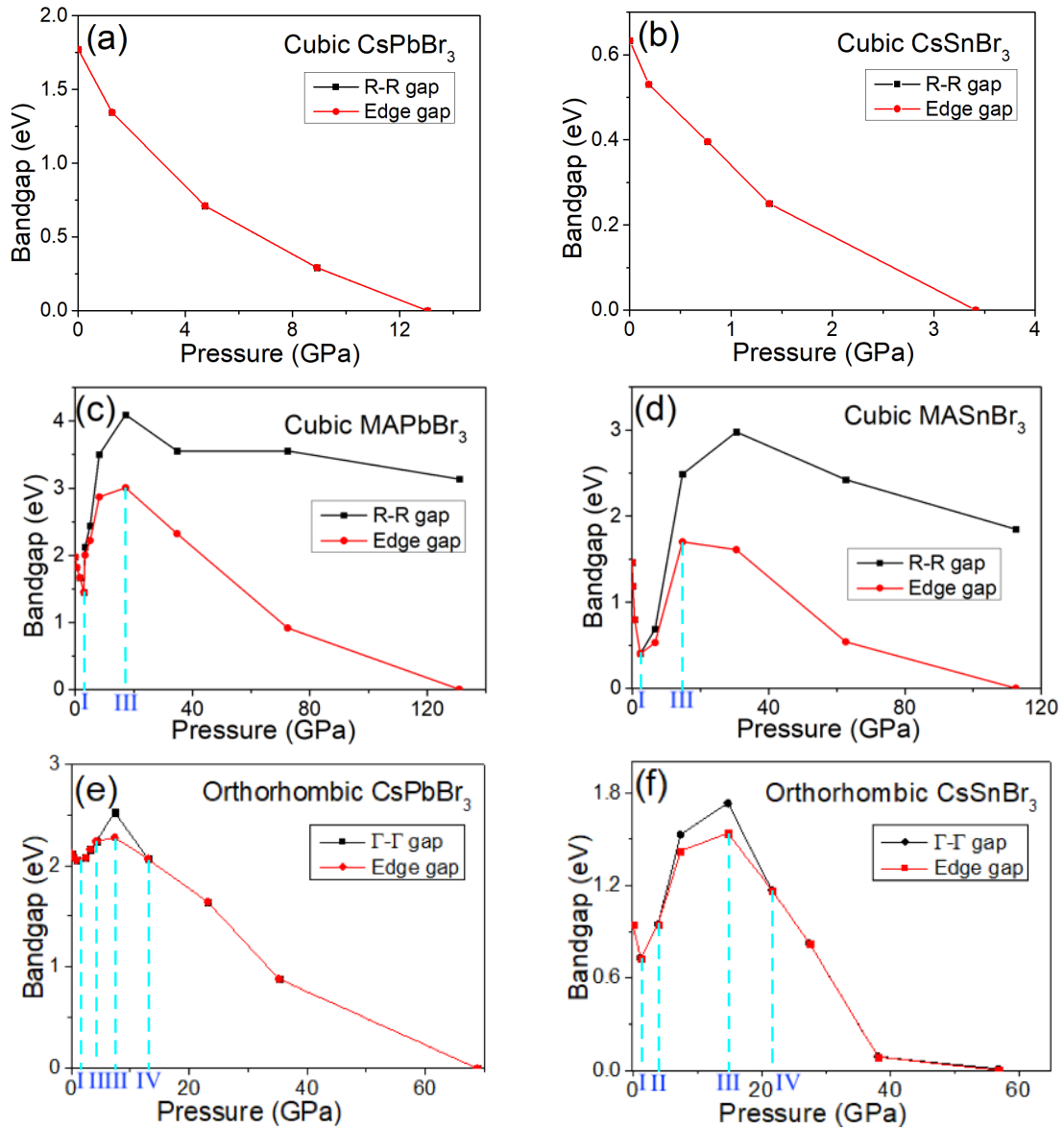

Figure 1: Band gap versus pressure for the six crystals: cubic MAPbBr$_3$, cubic MASnBr$_3$, cubic CsPbBr$_3$, cubic CsSnBr$_3$, orthorhombic CsPbBr$_3$ and orthorhombic CsSnBr$_3$. Both band gap between CBM and VBM and band gap at R-R (Γ-Γ) are presented.

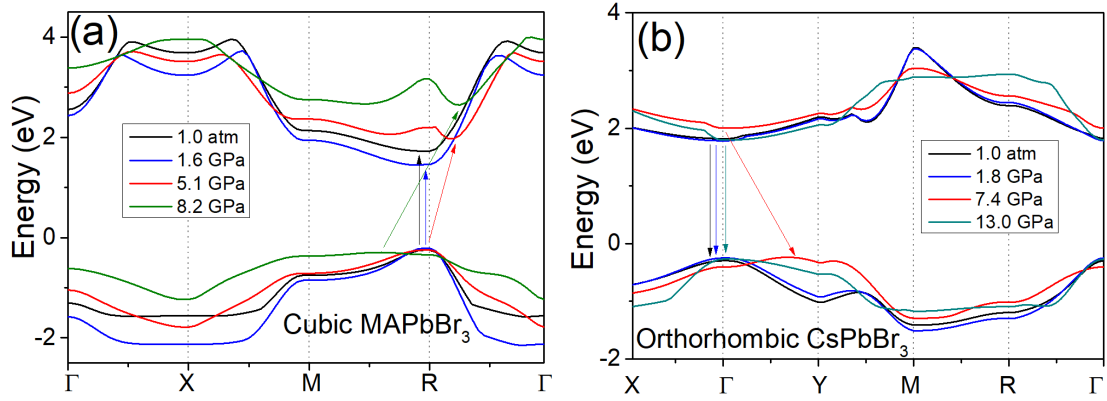

Figure 2: Band structure close to Fermi level of cubic MAPbBr$_3$ and orthorhombic CsPbBr$_3$ under four critical pressure values.

These electronic band gap properties variation of perovskites can be explained by pressure induced atomic structural change. In order to rationalize the direct-to-indirect band gap transition in cubic organic crystals at CP_I, direct-to-indirect and the inverse transitions in orthorhombic inorganic crystals at CP_II, CP_IV, respectively, crystal structures of cubic MAPbBr$_3$ (similar to MASnBr$_3$) and orthorhombic CsPbBr$_3$ (similar to CsSnBr$_3$) under pressures in the four critical stages (the same pressure values in Figure 2) are shown in Figure 3. Space groups of these eight crystal structures are determined using software Accelrys Materials Studio and shown below the figures in Figure 3. For cubic MAPbBr$_3$, the crystal is in cubic phase (Pm$\bar{3}$m) under ambient pressure, trigonal phase with symmetric mirror (R3m) at direct band gap state (1.6 GPa) before CP_I, low-symmetry trigonal phase (R3) at indirect band gap state (5.1, 8.2 GPa) after CP_I. For orthorhombic CsPbBr$_3$, the crystal maintains in orthorhombic phase (Pnma) under ambient pressure, at direct band gap state (1.8 GPa) before CP_II, and at indirect band gap state (7.4 GPa) between CP_II and CP_IV, with more and more severe octahedral rotation/distortion and resulting lower and lower structural symmetry sequentially. Crystal structure transforms to the high-symmetry hexagonal phase at direct band gap state (13 GPa) after CP_IV. Therefore, direct-to-indirect and indirect-to-direct transitions might be closely related to the reduction and enhancement of crystal symmetry, respectively. As for the magnitude evolution of band gap, before CP_III, lattice contraction that reduces bandgap and octahedral rotation which tends to enlarge bandgap compete with each other [30]. The relation between averaged angle Pb-Br-Pb that can describe octahedral rotation and averaged Pb-Br bond length which can evaluate lattice contraction magnitude of cubic MAPbBr$_3$ (similar to MASnBr$_3$) and orthorhombic CsPbBr$_3$ (similar to CsSnBr$_3$) are plotted in Figure 4. Corresponding bandgap-pressure relationships before CP_III are also plotted in Figure 4, amplifyingly. As can be noticed, for both organic and inorganic perovskite, abstract values of slopes increase suddenly with increasing pressure exactly at CP_I (4th data point for cubic MAPbBr$_3$ and 5th data point for orthorhombic CsPbBr$_3$). After CP_I, the effect of octahedral rotation exceeds lattice contraction and bandgap begins to go up.

In an electronic landscape, the phenomenon of band gap narrowing followed with band gap widening can be comprehensed as follow: for both organic perovskite and inorganic perovskite, only orbitals Pb (Sn) and Br contribute to CBM and VBM. VBM is antibonding organicization between Pb 6s and Br 4p orbitals. CBM state is governed by orbital interaction between Pb 6p and Br 4s orbitals. As can be identified in Figure 5 (a) (c), CBM is mostly a nonbonding localized state of Pb 6p orbital, thus not sensitive to pressure. Therefore, band gap is determined by the coupling between Pb 6s orbital and Br 4p orbital in VBM. Aforementioned Pb-Br bond length shortening facilitates the coupling effect, pushes up VBM and narrows the band gap. This explanation can also be applied to the two highly symmetric structures, cubic CsPbBr$_3$ and cubic CsSnBr$_3$. At CP_I, as shown in Figure 5 (e) (f), electron cloud distance between Pb 6s and Br 4p orbitals reaches to minimum and is ready to increase due to octahedral rotation evaluated by the aforementioned Pb-Br-Pb angle, so the coupling effect starts to become weaker, which leads to VBM drop and band gap widening. When pressure goes beyond CP_III, pressure induced metallization happens and band gap begins to narrow down all the way to zero when band cross occurs [31], which may be the result of strong coupling effect between Pb 6s and Br 4p orbitals in the small pressed crystal cells. Another way to explain this pressure induced metallization phenomenon is Goldhammer–Herzfeld criterion [32] which suggests that electrons can be ripped of the atoms or molecules with an infinitesimal perturbation originated from high crystal packing factor under high pressure. For the difference in CBM/VBM participation in direct-indirect mutual band gap transition between organic and inorganic perovskite, a possible explanation has been provided as follow: as shown in Figure 5 (a) (b), in organic perovskite, MA has polarity and orients along <111>, the same direction as Pb 6p orbital in CBM. This orientational match may induce a strong Coulomb interaction between MA and Pb, which can be identified from the morphology change of Pb 6p orbital under high pressure (figure 5 (b)) and may facilitate the CBM movement away from R point. The strong MA-Pb interaction may derive from the experimentally demonstrated organic molecule-to-octahedral interactions enhanced by pressure in organic halide perovskite [33]. While in inorganic perovskite (Figure 5 (c) (d)), no crystal orbital morphology change has been observed under high pressure (Figure 5 (d)), which may be subject to the non-polarity of Cs atom and the resulting small magnitude of Coulomb interaction between Cs and Pb that is too small to trigger CBM deviation. Compared to direct semiconductor, indirect semiconductor tends to have longer carrier lifetime, which is beneficial for photovoltage enhancement.

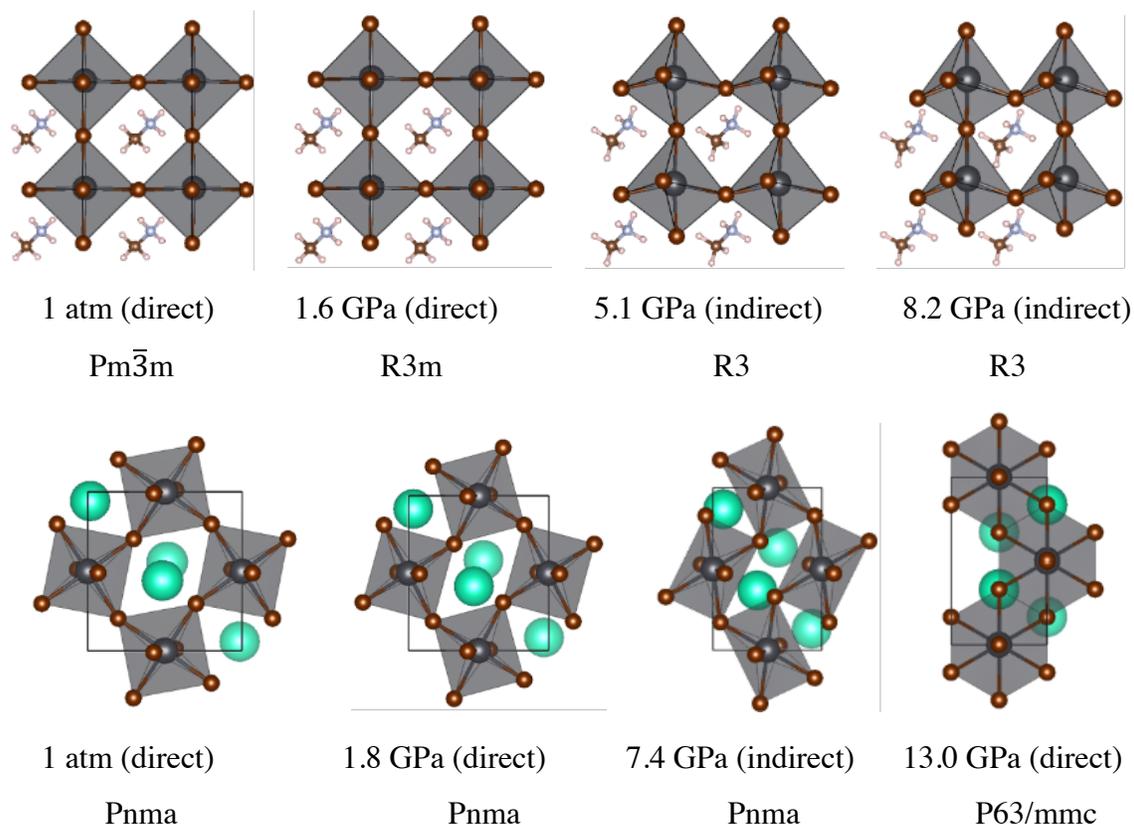

Figure 3: Up: MAPbBr$_3$ structures evolution under four critcal pressure values; down: CsPbBr$_3$ structures evolution under pressures. Corresponding pressure and space group symbols are shown below each structure.

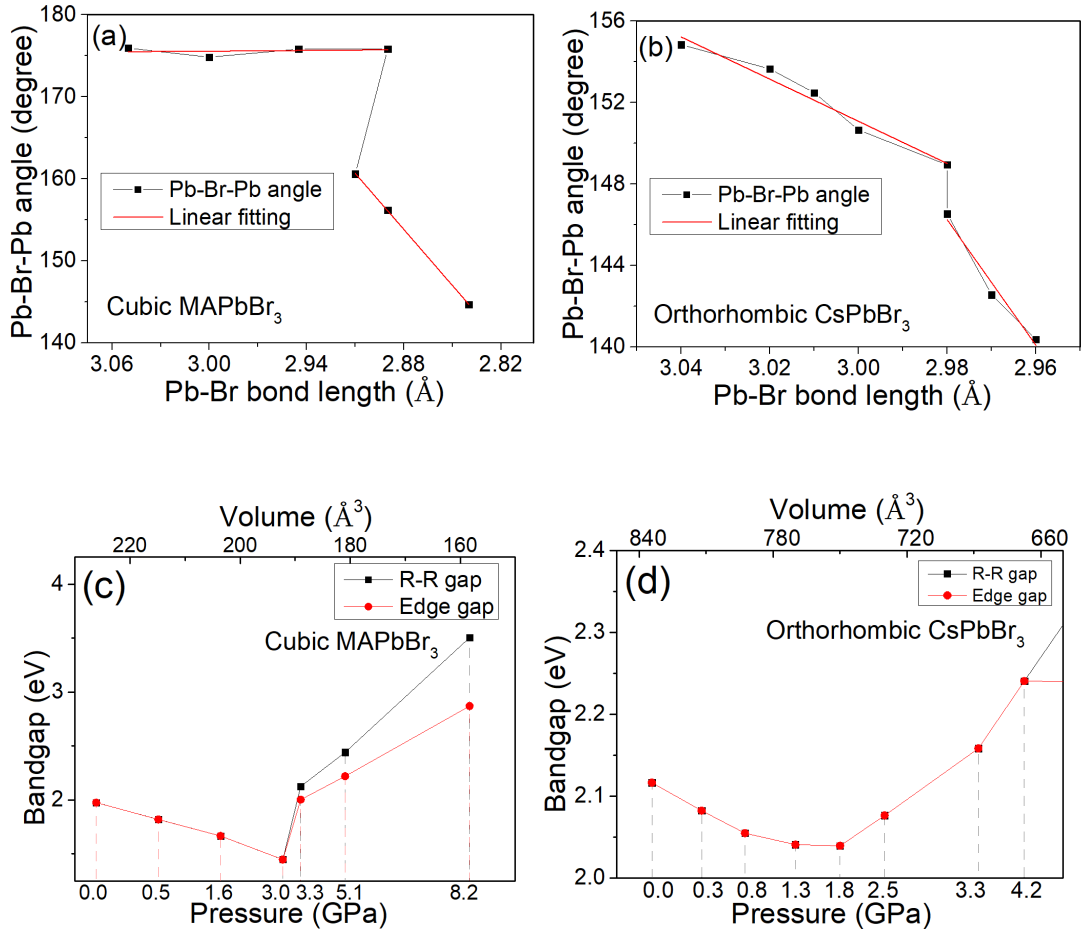

Figure 4: (a) (b): Averaged Pb-Br-Pb angle vs averaged Pb-Br bond length. Linear fittings are shown in red; (c) (d): amplified bandgap vs volume and pressure before CP_III. Data points are vertically matched between (a) and (c), (b) and (d), respectively.

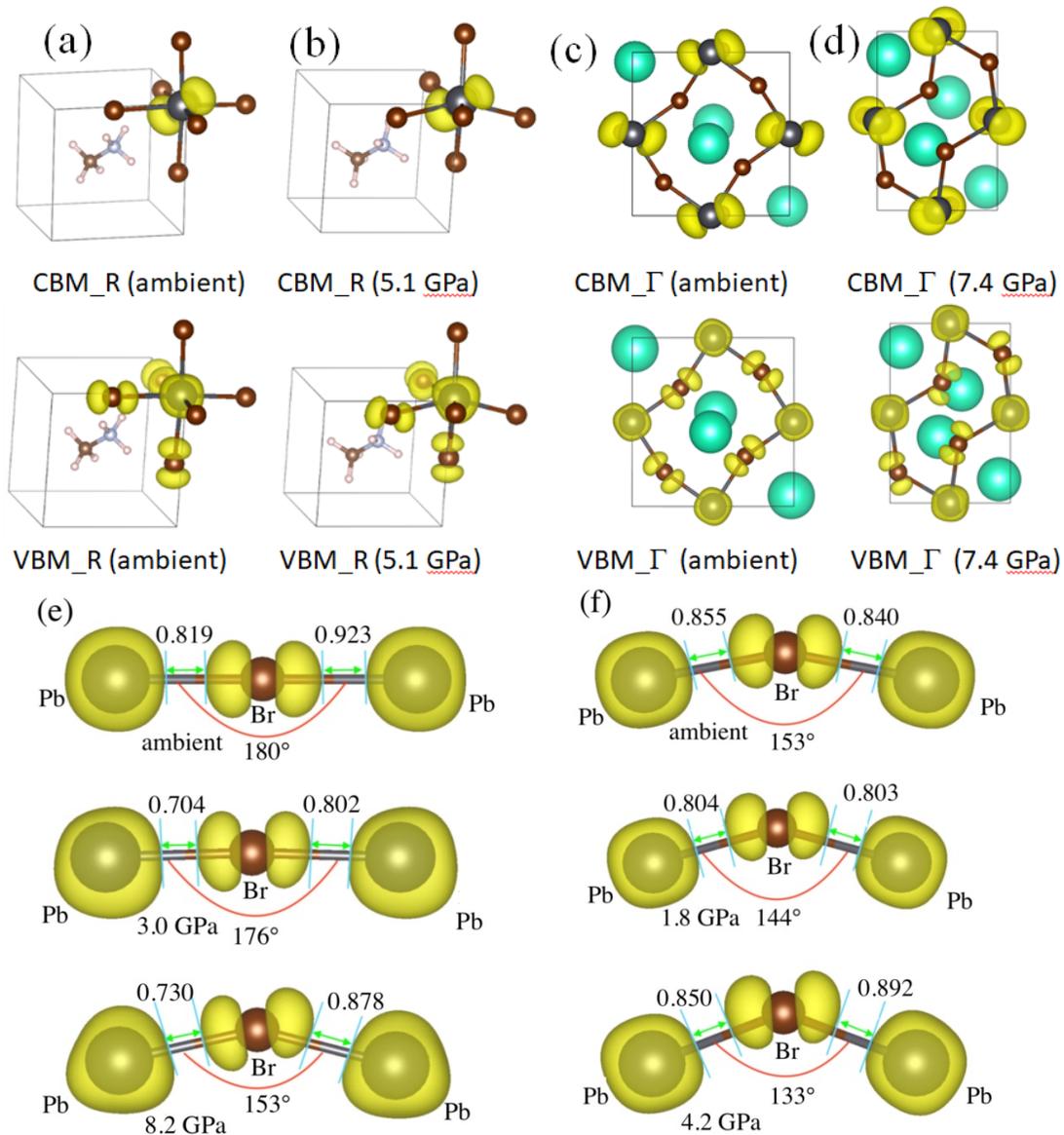

Figure 5: (a) (b) CBM and VBM electron density character of cubic MAPbBr$_3$ under ambient pressure and after CP_I, respectively. (c) (d): CBM and VBM electron density character of orthorhombic CsPbBr$_3$ under ambient pressure and after CP_I, respectively. (e) (f): Bond angles and cloud distances for Pb-Br-Pb of cubic MAPbBr$_3$ and orthorhombic CsPbBr$_3$. Electron cloud distances decrease and then increase with increased pressure, corresponding to narrowing and widening of band gap.

## Conclusions

In summary, with a wide range of pressure being applied, band structure evolution of organic and inorganic perovskites with different structural symmetry have been investigated and rationalized. Both the magnitude and direct-indirect nature of band gap can be tuned with the help of pressure, which implies a green and facile way for engineering optoelectronic materials. In addition, from a scientific view, pressure enhanced structural symmetry alternation is highly possible for triggering direct-indirect mutual transition of band gap and unidentical CBM/VBM behaviors in organic and inorganic perovskites may be subject to different molecular polarity outside octahedrals. It is believed that the whole picture formed in this work on perovskite under pressure effects provides useful implications for further design and property enhancement of perovskite materials.

**Acknowledgements**

This work was jointly supported by National Natural Science Foundation of China (Grant No. 11504332), and Outstanding Young Talent Research Fund of Zhengzhou University (Grant No. 1521317008). The calculations were performed on the high performance computational center of Zhengzhou University.